\documentclass[twocolumn,aps,superscriptaddress]{revtex4}
\usepackage[utf8]{inputenc}
\usepackage{bbm}
\usepackage{amssymb}
\usepackage{amsmath}
\usepackage{graphicx}
\usepackage[normalem]{ulem}
\usepackage[dvips]{color}
\usepackage{appendix}
\usepackage{color}

\def\es0{$E_{\rm sym}(\rho_0)$}

\def\us0{$U_{\rm sym}(\rho_0,k_F)$~}

\def\l0{$L(\rho_0)$~}

\newcommand{\beq}{\begin{equation}}
\newcommand{\eeq}{\end{equation}}
\newcommand{\ba}{\begin{array}}
\newcommand{\ea}{\end{array}}
\newcommand{\bea}{\begin{eqnarray}}
\newcommand{\eea}{\end{eqnarray}}
\newcommand{\bi}{\begin{itemize}}  
\newcommand{\ei}{\end{itemize}}
\newcommand{\ben}{\begin{enumerate}} 
\newcommand{\een}{\end{enumerate}}
\newcommand{\bc}{\begin{center}}
\newcommand{\ec}{\end{center}}

\newcommand{\cQM}{{c^{\phantom{1}}_{\rm QM}}}

\newcommand{\cQMsq}{c^2_{\rm QM}}

\newcommand{\De}{\Delta}
\newcommand{\ep}{\varepsilon}

\begin{document}

\title{Bayesian inference of dense matter EOS encapsulating a first-order hadron-quark phase transition from observables of canonical neutron stars}

\author{Wen-Jie Xie}
\affiliation{Department of Physics, Yuncheng University, Yuncheng 044000, China}
\author{Bao-An Li\footnote{Corresponding author: Bao-An.Li@tamuc.edu}}
\affiliation{Department of Physics and Astronomy, Texas A$\&$M University-Commerce, Commerce, TX 75429, USA}
\date{\today}

\begin{abstract}

\begin{description}
\item[Background:]The remarkable progress in recent multimessenger observations of both isolated neutron stars (NSs) and their mergers has provided some of the much needed data to improve our understanding about the Equation of State (EOS) of dense neutron-rich matter. Various EOSs with or without some kinds of phase transitions from hadronic to quark matter (QM) have been widely used in many forward-modelings of NS properties. Direct comparisons of these predictions with observational data sometimes also using $\chi^2$ minimizations have provided very useful constraints on the model EOSs. However, it is normally difficult to perform uncertain quantifications and analyze correlations of the EOS model parameters involved in forward-modelings especially when the available data are still very limited.

\item[Purpose:] We infer the posterior probability distribution functions (PDFs) and correlations of nine parameters characterizing the EOS of dense neutron-rich matter encapsulating a first-order hadron-quark phase transition from the radius data of canonical NSs reported by LIGO/VIRGO, NICER and Chandra Collaborations. We also infer the QM mass fraction and its radius in a 1.4 M$_{\odot}$ NS and predict their values in more massive NSs.

\item[Method:] Meta-modelings are used to generate both hadronic and QM EOSs in the Markov-Chain Monte Carlo sampling process within the Bayesian statistical framework. An explicitly isospin-dependent parametric EOS for the $npe\mu$ matter in NSs at $\beta$ equilibrium is connected through the Maxwell construction to the QM EOS described by the constant speed of sound (CSS) model of Alford, Han and Prakash with and without using the Seidov stability condition for first-order phase transitions.

\item[Results:] In the default calculation with the Seidov stability condition, we find that (1) The most probable values of the hadron-quark transition density $\rho_t/\rho_0$ and the relative energy density jump there $\De\ep/\ep_t$ are $\rho_t/\rho_0=1.6^{+1.2}_{-0.4}$ and $\De\ep/\ep_t=0.4^{+0.20}_{-0.15}$ at 68\% confidence level, respectively. The corresponding probability distribution of QM fraction in a 1.4 M$_{\odot}$ NS peaks around 0.9 in a 10 km sphere.
Strongly correlated to the PDFs of $\rho_t$ and $\De\ep/\ep_t$, the PDF of the QM speed of sound squared $\cQMsq/c^2$ peaks at $0.95^{+0.05}_{-0.35}$, and the total probability of being less than 1/3 is very small. (2) The correlations between PDFs of hadronic and QM EOS parameters are very weak. While the most probable values of parameters describing the EOS of symmetric nuclear matter remain almost unchanged, the high-density symmetry energy parameters of neutron-rich matter are significant different with or without considering the hadron-quark phase transition. Removing the Seidov condition, while there are appreciable and interesting changes in the PDFs of quark matter EOS parameters, the qualitative conclusions remain the same.

\item[Conclusions:] The available astrophysical data considered together with all known EOS constraints from theories and terrestrial nuclear experiments prefer the formation of a large volume of QM even in canonical NSs.

\end{description}
\end{abstract}

\maketitle

\section{Introduction}

 Probing the Equation of State (EOS) of dense neutron-rich matter has been a long standing and shared goal of both astrophysics and nuclear physics. Much progress has been made in realizing this goal using various messengers from both isolated neutron stars (NSs) and their mergers especially since LIGO/VIRGO's observation of GW170817. For recent reviews, see, e.g., Refs. \cite{Baiotti,BALI19,Const,David,Capano20,Kat20,AngLi}. Among the many interesting questions studied in the literature, significant efforts have been devoted for a long time to investigating whether quark matter exists in NSs, the nature as well as where and when the hadron-quark phase transition may happen, what its size and EOS may be if quark matter does exist in NSs or can be created during their mergers, see, e.g.,
 Refs. \cite{Seidov:1971,Haensel:1983,Haensel:1987,Lindblom,Glen,Web,Andreas,Mark,Han-Xiamen,Frank1,GSI1,Frank2,GSI2}. Since the earlier debate \cite{Ozel,Alford} on whether the mass and radius of EXO 0748-676 can rule out the existence of its quark core, despite of the great progresses made using various astrophysical data including the latest ones from LIGO/VIRGO, NICER and Chandra observations, state-of-the-art theories and models as well as updated nuclear physics constraints, see, e.g., Refs. \cite{David,Ann20,Con2,Chr,Tang} and references therein, no consensus has been reached on most of the issues regarding the nature and EOS of dense NS matter.

Most of the studies about the nature of hadron-quark phase transition and the size of possible quark matter core in NSs have been carried out by using the traditional forward-modeling approach based on various theories for both the hadronic and quark phases, perhaps except very few recent studies using Bayesian analyses, see, e.g., Refs. \cite{David,Tang}.
Often, various forms of polytropes or spectrum functions are used to interpolate the NS EOS starting slightly above the saturation density $\rho_0$ of nuclear matter (below which reliable theoretical predictions and some experimental constraints exist) to very high densities where predictions of perturbative QCD exist. Comparisons of model predictions with observational data have provided very useful constraints on the model EOSs considered. Although $\chi^2$ minimizations are sometimes used, often conclusions are strongly model dependent. Moreover, it is normally difficult to perform uncertain quantifications and analyze correlations of the EOS model parameters involved in forward-modelings especially when the available data are still very limited.

In this work, meta-modelings are used for both hadronic and quark phases to construct very generally the EOSs of NS matter. An explicitly isospin-dependent EOS \cite{Zhang2018} for the $npe\mu$ matter in NS at $\beta$ equilibrium is connected through the Maxwell construction to the constant speed of sound (CSS) quark matter EOS \cite{CSS1}. With totally 9 parameters in their prior ranges allowed by general physical principles and available constraints, the constructed NS EOS is so generic that it can essentially mimic any NS EOS available in the literature. Without restrictions and possible biases of underlying energy density functionals of specific theories, we infer the probability distribution functions (PDFs) of the nine EOS parameters using the available NS radius data from LIGO/VIRGO, NICER and Chandra, satisfying the causality and dynamical stability condition within the Bayesian statistical framework.
We found that the available astrophysical data considered together with all known EOS constraints from theories and terrestrial nuclear experiments prefer the formation of a large volume of QM even in canonical NSs.

\section{Theoretical approach}\label{theory}
Here we summarize the major features of our approach. In the CSS model of Alford, Han and Prakash \cite{CSS1} for hybrid NSs,  the pressure in NSs is parameterized as

\beq
\ep(p) = \left\{\!
\begin{array}{ll}
\ep_{\rm HM}(p) & p<p_{t} \\
\ep_{\rm HM}(p_{t})+\De\ep+c_{\rm QM}^{-2} (p-p_{t}) & p>p_{t}
\end{array}
\right.\ ,
\label{eqn:EoSqm1}
\eeq
where $\ep_{\rm HM}(p)$ is the hadronic matter (HM) EOS below the hadron-quark transition pressure $p_{t}$, $\De\ep$ is the discontinuity in energy density $\ep$ at the transition,
and $\cQM$ is the QM speed of sound. Once the HM EOS $\ep_{\rm HM}(p)$ is specified, the transition pressure $p_{t}$ and energy density $\ep_t$ are uniquely related to the hadron-quark transition density $\rho_t$. In our Bayesian analyses using the CSS model, we use the $\rho_t/\rho_0$, $\De\ep/\ep_t$ and $\cQMsq/c^2$ as three independent parameters to be generated randomly with uniform prior PDFs in the range of 1 to 6 (or 10 for comparison), $0.2-1$ and $0-1$, respectively. We thus use the CSS model as a generic meta-model for generating the QM EOS.

In several recent applications of the CSS model, see, e.g., Refs. \cite{Dri,Han-apj,Miao,Kat}, various HM EOSs predicted by microscopic nuclear many-body theories and/or phenomenological models have been used. These HM EOSs are often restricted by the underlying energy density functionals of the theories used and are usually not flexible enough in statistical analyses to explore the whole EOS parameter space permitted by general physics principles and known constraints as pointed out already in Refs. \cite{Con2,Tsang}. On equal footing as the generic QM EOS, we use the meta-model of Ref. \cite{Zhang2018} for generating the HM EOS.  The explicitly isospin dependence of the latter built into the EOS at the level of average nucleon energy in neutron-rich matter is an important distinction compared to directly parameterizing the HM pressure as a function of energy or baryon density with piecewise polytropes or spectrum functions. Such kinds of parameterizations with minor variations for HM EOSs have been widely used in both nuclear physics, see, e.g. Refs. \cite{MM1,MM2} and astrophysics applications, see, e.g., Refs. \cite{BALI19,Con2,Zhang2018,Tsang,Zhang19apj,Zhang19EPJA,Zhang20a,France1,Xie19,Xie20a,France2,India}. For this work, we calculate the pressure within the $npe\mu$ model for the core of NSs using
\begin{equation}\label{pressure}
  P(\rho, \delta)=\rho^2\frac{d\epsilon_{\rm HM}(\rho,\delta)/\rho}{d\rho}
\end{equation}
where the HM energy density $\epsilon_{\rm HM}(\rho, \delta)=\epsilon_n(\rho, \delta)+\epsilon_l(\rho, \delta)$ with $\epsilon_n(\rho, \delta)$ and $\epsilon_l(\rho, \delta)$ being the energy densities of nucleons and leptons, respectively.
While the $\epsilon_l(\rho, \delta)$ is calculated using the noninteracting Fermi gas model \citep{Oppenheimer39}, the $\epsilon_n(\rho, \delta)$ is from
\begin{equation}\label{lepton-density}
  \epsilon_n(\rho, \delta)=\rho [E(\rho,\delta)+M_N]
\end{equation}
where $M_N$ is the average nucleon mass. The average energy per nucleon $E(\rho,\delta)$ in neutron-rich matter of isospin asymmetry $\delta=(\rho_n-\rho_p)/\rho$ is parameterized in terms of the energy per nucleon $E_0(\rho)\equiv E(\rho ,\delta=0)$ in symmetric nuclear matter (SNM) and the symmetry energy $E_{\rm{sym}}(\rho )$ as
\cite{Bom91}
\begin{equation}
E(\rho,\delta)=E_0(\rho)+E_{\rm{sym}}(\rho)\delta^{2}.
\end{equation}
The $E_0(\rho)$ and $E_{\rm{sym}}(\rho )$ are parameterized respectively as
\begin{eqnarray}\label{E0para}
E_{0}(\rho)&=&E_0(\rho_0)+\frac{K_0}{2}(\frac{\rho-\rho_0}{3\rho_0})^2+\frac{J_0}{6}(\frac{\rho-\rho_0}{3\rho_0})^3,\\
E_{\rm{sym}}(\rho)&=&E_{\rm{sym}}(\rho_0)+L(\frac{\rho-\rho_0}{3\rho_0})+\frac{K_{\rm{sym}}}{2}(\frac{\rho-\rho_0}{3\rho_0})^2\nonumber\\
&+&\frac{J_{\rm{sym}}}{6}(\frac{\rho-\rho_0}{3\rho_0})^3\label{Esympara}
\end{eqnarray}
where $E_0(\rho_0)=-15.9$ MeV. Guided by our prior knowledge from both astrophysics and nuclear physics, see, e.g., Ref. \cite{BALI19} for a recent review, we generate randomly with uniform prior PDFs for the six HM EOS parameters $K_0$, $J_0$, $E_{\rm{sym}}(\rho_0)$, L, $K_{\rm sym}$ and $J_{\rm sym}$ in their currently known uncertain ranges: $220\leq K_0 \leq 260$ MeV,  $-800 \leq J_{0}\leq 400$ MeV,  $28 \leq E_{\rm sym}(\rho_0) \leq 36$ MeV,  $30 \leq L \leq 90$ MeV, $-400 \leq K_{\rm{sym}} \leq 100$ MeV, and $-200 \leq J_{\rm{sym}}\leq 800$ MeV, respectively.

The density profile of isospin asymmetry $\delta(\rho)$ in charge neutral NSs at $\beta$ equilibrium is uniquely determined by the symmetry energy $E_{\rm{sym}}(\rho)$. Once the $\delta(\rho)$ is determined, both the
$P(\rho, \delta)$ and $\epsilon_{\rm HM}(\rho,\delta)$ become barotropic functions of density $\rho$. The core EOS outlined above is then connected smoothly to the NV EOS \cite{Negele73} for the inner crust and the BPS EOS \cite{Baym1971} for the outer crust using the crust-core transition density and pressure evaluated consistently using a thermodynamical approach from the core side with the same parameters given above \cite{Zhang2018}.

The role of Seidov stability condition for first-order phase transitions \cite{Seidov:1971,Haensel:1983,Lindblom}
\beq
\frac{\De\ep}{\varepsilon_{t}} \leq \frac{1}{2} + \frac{3}{2}\frac{p_{t}}{\varepsilon_{t}}
\label{eqn:stability}
\eeq
in forming different topologies of hybrid stars was studied in detail in Ref. \cite{CSS1}. Enforcing the above condition leads to a stable connected hybrid branch on the mass-radius curve but does not rule out a disconnected stable hybrid branch. On the other hand, without the above condition, additional hadronic or a hadronic together with a disconnected hybrid branch may also be formed depending on the hadron-quark transition density and the energy jump there. Overall, it was concluded that the Seidov condition is not a good guide for the presence of observable hybrid branches \cite{CSS1}. In this work, we carried out two calculations. In the default calculation to be presented first in the following, we enforce the Seidov condition. We focus on discussing if quark matter may exist and how big it may be in canonical neutron stars of mass 1.4 M$_{\odot}$ as well as if the inferred EOS parameters are consistent with their known constraints from both astrophysics and nuclear physics. In another calculation, we remove the Seidov stability condition and then compare the resulting PDFs of the CSS model parameters with those from the default calculation. A comparison of these two calculations will allow us to evaluate the role of the Seidov condition on the inferred PDFs of quark matter EOS parameters.

\begin{figure*}[htb]
\begin{center}
\resizebox{0.9\textwidth}{!}{
 \includegraphics[width=\linewidth]{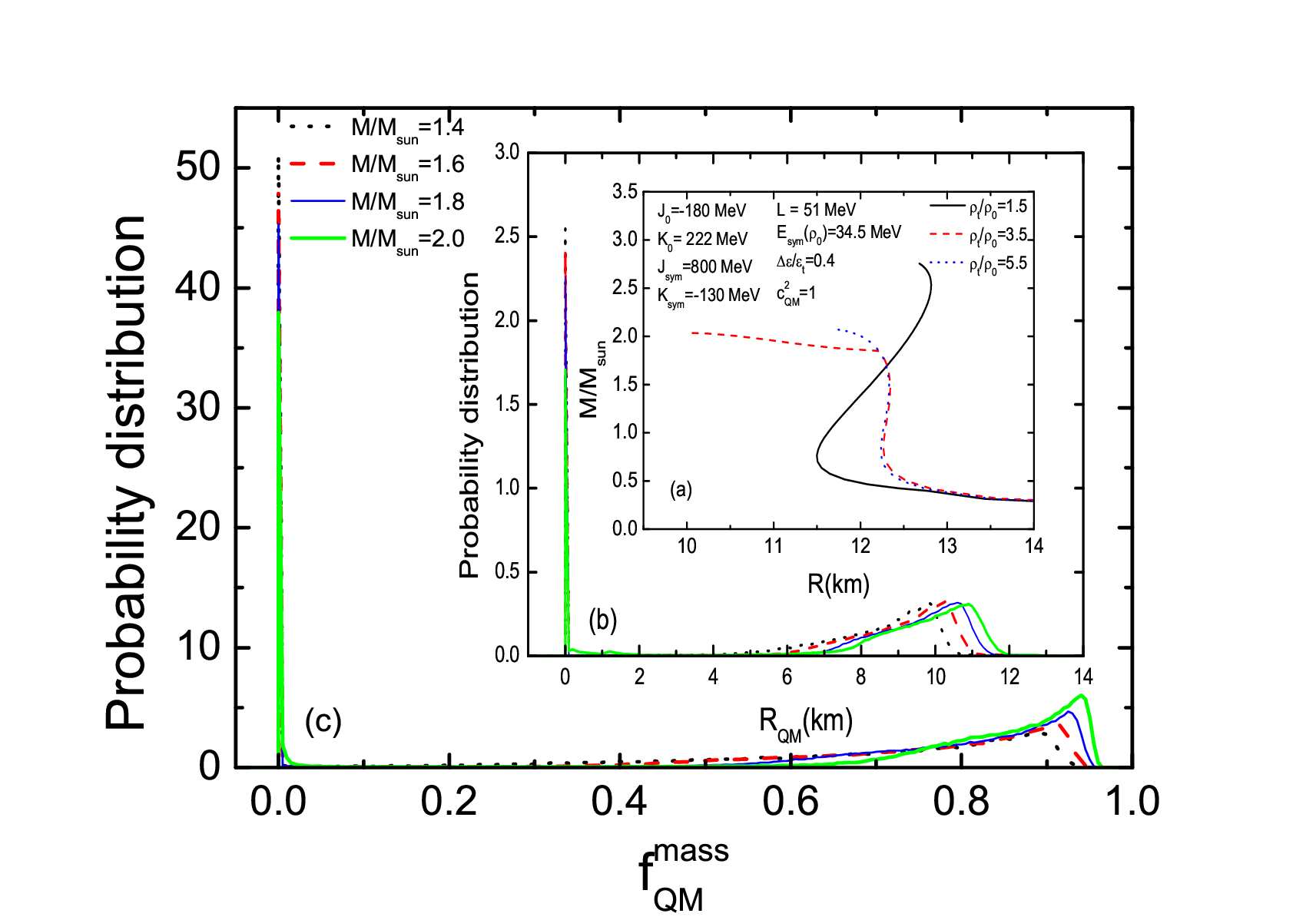}
 }
  \caption{(Color online) The inner window: mass-radius sequences of selected samples with the three different hadron-quark transition densities but all other parameters fixed at the values specified. The middle and outer windows are the normalized probability distribution of the quark matter radius and fraction, respectively, from all accepted EOSs in the Bayesian analysis.}\label{QM-fr}
\end{center}
\end{figure*}
\begin{figure*}[htb]
\begin{center}
 \resizebox{0.9\textwidth}{!}{
  \includegraphics[scale=1]{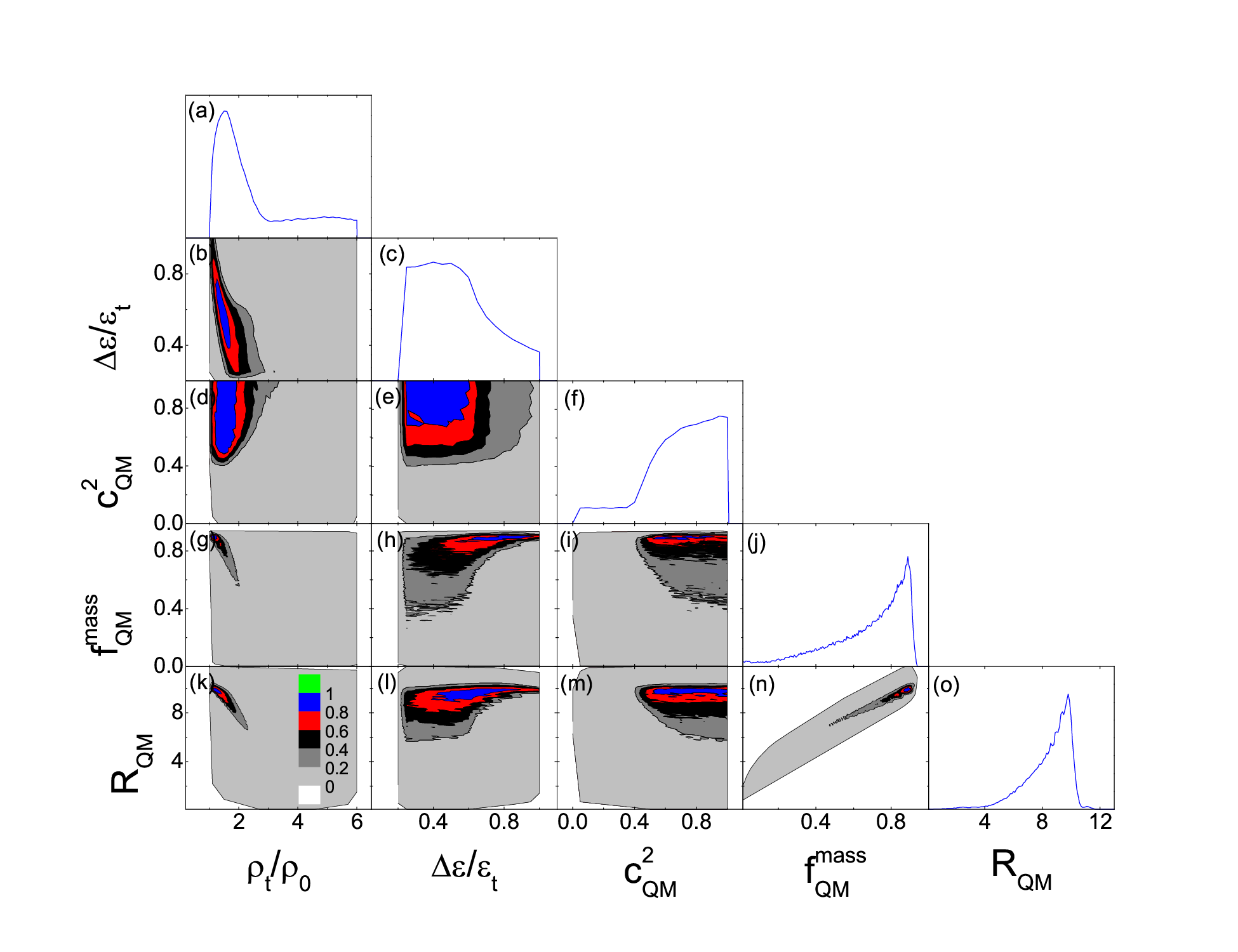}}
\vspace{-0.5cm}
  \caption{(Color online) The posterior probability distribution functions (in arbitrary units) and correlations of the three quark matter EOS parameters as well as the fraction and radius of quark matter in hybrid neutron stars of mass 1.4 M$_{\odot}$.}\label{QM}
\end{center}
\end{figure*}

We use the standard Bayesian formalism and the Markov-Chain Monte Carlo (MCMC) technique to evaluate the posterior PDFs of EOS parameters as in our previous work where no hadron-quark phase transition was
considered but using the same HM meta-model for the entire core of NSs \cite{Xie19,Xie20a}. For easy of the following discussions, we notice the following key inputs and aspects of our Bayesian analyses:
\begin{itemize}
\item The likelihood function $P[D|{\cal M}(p_{1,2,\cdots 9})]$ measures the ability of the model $\cal M$ with 9 parameters $p_{1,2,\cdots 9}$ to reproduce the observational data D. We use \cite{Xie19,Xie20a}
\begin{equation}\label{Likelihood}
  P[D|{\cal M}(p_{1,2,\cdots 9})]=P_{\rm{filter}} \times P_{\rm{mass,max}} \times P_{\rm{radius}} \nonumber
\end{equation}
where the $P_{\rm{filter}}$ is a filter selecting EOS parameter sets satisfying the following conditions: (i) The crust-core transition pressure always stays positive; (ii) At all densities, the thermaldynamical stability condition (i.e., $dP/d\varepsilon\geq0$), the Seidov stability of Eq. (\ref{eqn:stability}), the causality condition (i.e, the speed of sound is always less than that of light) are satisfied. The $P_{\rm{mass,max}}$ stands for the requirement that each accepted EOS has to be stiff enough to support the observed NS maximum mass $M_{\rm{max}}$. We present results with $M_{\rm{max}}$=1.97 M$_{\odot}$ to be consistent with that used by the LIGO/VIRGO Collaborations
in their extraction of the NS radius from GW170817 \cite{LIGO18}. Using 2.01 or 2.14 M$_{\odot}$ for $M_{\rm{max}}$ has only some minor quantitative effects.

\item
We use the following radii of canonical NSs as independent data: 1) $R_{1.4}=11.9\pm 1.4$ km extracted by the LIGO/VIRGO Collaborations from GW170817 \cite{LIGO18}, 2) $R_{1.4}=10.8^{+2.1}_{-1.6}$ extracted independently also from GW170817 by De {\it et al.} \cite{De18}, 3) $R_{1.4}=11.7^{+1.1}_{-1.1}$ from earlier analysis of quiescent low-mass X-ray binaries observed by Chandra and XMM-Newton observatories \cite{Lattimer14}, and 4)
$R=13.02^{+1.24}_{-1.06}$ km with mass $M=1.44^{+0.15}_{-0.14}$ M$_{\odot}$ \cite{Miller19} or $R=12.71^{+1.83}_{-1.85}$ km with mass $M=1.34\pm0.24$ M$_{\odot}$ \cite{Riley19} for PSR J0030+0451
from NICER Collaboration. The errors quoted are at 90\% confidence level. Correspondingly, the $P_{\rm radius}$ is a product of four Gaussian functions, i.e.,
\begin{equation}\label{Likelihood-radius}
  P_{\rm radius}=\prod_{j=1}^{4}\frac{1}{\sqrt{2\pi}\sigma_{\mathrm{obs},j}}\exp[-\frac{(R_{\mathrm{th},j}-R_{\mathrm{obs},j})^{2}}{2\sigma_{\mathrm{obs},j}^{2}}]\nonumber
\end{equation}
where $\sigma_{\mathrm{obs},j}$ represents the $1\sigma$ error bar of the radius from the observation $j$ while $R_{\mathrm{th},j}$ is the corresponding theoretical prediction.  More details can be found in our previous work in Refs. \cite{Xie19,Xie20a}.

\item
In the MCMC process of sampling the posterior PDFs of EOS parameters, we throw away the initial 100,000 burn-in steps/EOSs before the stationary state is reached. Afterwards, we generate 1600,000 steps/EOSs to calculate the posterior PDFs and correlations of EOS parameters. The acceptance rate is about 15\%.

\end{itemize}

We emphasize that a fundamental assumption made in the CSS model is that once the energy density reached in the core of NS is higher than a critical value $\ep_c=\ep_{\rm HM}(p_{t})+\De\ep$, QM will be formed through the first-order hadron-quark phase transition. All results presented here thus have to be understood with this assumption in mind.

\section{Results and discussions}
In the following, we first present results of the default calculation with the Seidov condition. Effects of the latter on the PDFs of quark matter EOS parameters will be examined in Section \ref{Se} by comparing the default calculation with a calculation without using the Seidov condition.

\subsection{Quark matter fraction and size in hybrid stars}
Shown in Fig. \ref{QM-fr} (a) are the mass-radius sequences in selected samples with the hadron-quark transition density $\rho_t/\rho_0=1.5,3.5$ and 5.5, respectively, while all other parameters are fixed at the values specified in the figure (notice in particular that $\cQMsq/c^2=1$.) As expected, the stable hybrid branches are all connected to the NS branches.

In the study of hybrid stars, a key question has been whether the densities reached inside NSs are high enough to form a sizeable QM core. To answer this question, we show in Fig. \ref{QM-fr} (c) and (b) the normalized probability distribution of the QM fraction $f^{\rm mass}_{\rm QM}$ (defined as the ratio of QM mass over the total NS mass) and the corresponding QM radius $R_{\rm QM}$,  in regions where the energy density is higher than the QM critical energy density $\ep_c$ in NSs of mass 1.4, 1.6, 1.8 and 2.0 M$_{\odot}$, respectively, in the default Bayesian analysis with the nine EOS parameters. It is interesting to see the two peaks indicating the formation of purely hadronic and hybrid stars. The major peaks at $f^{\rm mass}_{\rm QM}=0$ correspond to pure hadronic NSs in cases where the $\ep_c$ is always higher than the maximum energy density at the core of the NSs considered. The second peaks around $f^{\rm mass}_{\rm QM}= 0.90\sim 0.95$ and $R_{\rm QM}=10\sim 11$ km corresponds to the formation hybrid stars consisting of mostly quark matter. While the probability ratio of the two peaks is about 6.7, the total probability of forming hybrid stars with $f^{\rm mass}_{\rm QM}$ higher than 0.1 is  77.6\%, 81.8\%, 85.2\% and 88.7\% for M=1.4, 1.6, 1.8 and 2.0 M$_{\odot}$, respectively.

By changing the prior range of hadron-quark transition density $\rho_t/\rho_0$ from the default 1-6 to 1-10, we found very little effect. We also found that correlations between the HM and QM EOS parameters are very weak, thus in the following we present the PDFs and correlations of quark matter and hadronic matter EOSs separately.

\subsection{Posterior probability distribution functions of quark matter EOS parameters and their correlations}
Shown in Fig. \ref{QM} are the posterior PDFs and correlations of QM EOS parameters $\rho_t/\rho_0$, $\De\ep/\ep_t$ and $\cQMsq/c^2$, as well as the $f^{\rm mass}_{\rm QM}$ and $R_{\rm QM}$ for canonical NSs in the default calculation. Several interesting features deserve emphasizing:
\begin{itemize}
\item The most probable values of the QM EOS parameters are $\rho_t/\rho_0=1.6^{+1.2}_{-0.4}$,
$\De\ep/\ep_t=0.4^{+0.20}_{-0.15}$ and $\cQMsq/c^2=0.95^{+0.05}_{-0.35}$ at 68\% confidence level.
Because the transition density peaks at a rather low density, and the energy jump at the transition is also relatively low, the QM stiffness
represented by its $\cQMsq$ value is rather high to provide the necessary pressure in QM. Since the average baryon density of a canonical NS with a 12 km radius is about $2\rho_0$, it is thus not surprising that for canonical NSs the PDFs of QM fraction and its radius peak around $f^{\rm mass}_{\rm QM}\approx 0.9$ and $R_{\rm QM}\approx 10$ km, respectively.

\item The total probability for $\cQMsq/c^2\leq 1/3$ is rather small. The considered astrophysical data informed us clearly that the value of $\cQMsq/c^2$ in QM is likely very high while the strength of the phase transition measured
with the energy density jump $\De\ep/\ep_t$ is modest (around 0.4).

\item The $f^{\rm mass}_{\rm QM}$, $R_{\rm QM}$ and $\De\ep/\ep_t$ are all anti-correlated with $\rho_t/\rho_0$ as one expects. When the transition density is low and the energy jump is weak, the required $\cQMsq/c^2$ has an approximately equally high probability to be between 0.5 to 1.

\end{itemize}

\subsection{The role of the speed of sound in quark matter}
Motivated by perturbative QCD predictions at extremely high densities or the casual limit, often in forward-model predictions one sets $\cQMsq/c^2$=1/3 or 1 among other constants examined. In fact, much efforts have been devoted to finding signatures/imprints of $\cQMsq/c^2$ from/on astrophysical observables especially since LIGO/VIRGO Collaborations' recent discovery that GW190814's secondary component has a mass of (2.50-2.67) M$_{\odot}$, see, e.g., Ref. \cite{Tan} and references therein.

\begin{figure}[htb]
\begin{center}
 \resizebox{0.5\textwidth}{!}{
 \includegraphics[width=\linewidth]{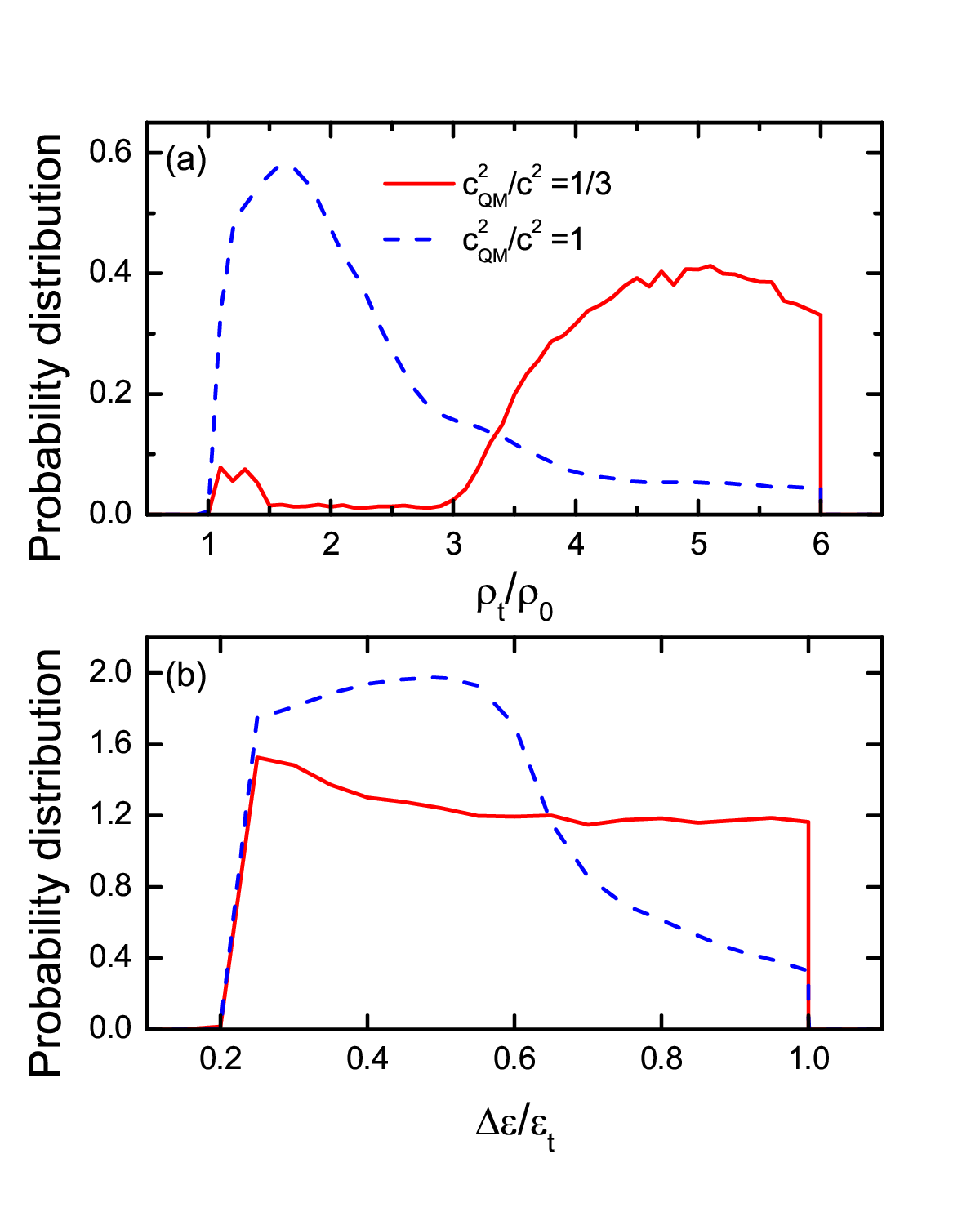}
 }
  \caption{(Color online) The posterior probability distribution functions of the hadron-quark matter transition density (upper) and the energy density jump at the transition (lower) from Bayesian analyses by setting the quark matter speed of sound squared $\cQMsq/c^2$ to 1/3 (solid curves) and 1 (dashed curves), respectively.}\label{CS}
\end{center}
\end{figure}

While in our default Bayesian analysis we have generated $\cQMsq/c^2$ randomly with a uniform prior PDF in the range of 0 to 1, it is interesting to compare the
default results with calculations setting $\cQMsq/c^2$ to certain constants. Shown in Fig. \ref{CS} are the posterior PDFs of the transition density (upper) and the jump of energy density there (lower) with $\cQMsq/c^2$=1/3 and 1, respectively.  While the results with $\cQMsq/c^2=1$ are very close to the default ones, setting $\cQMsq/c^2$=1/3 requires a much higher transition density and a larger energy density jump. This is simply because the resulting very soft QM EOS can't support the NSs considered if the hadron-quark transition happens at too low densities. Consequently, only very small QM fractions are allowed in the hybrid NSs. Quantitatively, we find that with $\cQMsq/c^2$=1/3 the $f^{\rm mass}_{\rm QM}$ has a value of only 2.3\%, 2.3\%, 2.3\% and 2.8\% for 1.4, 1.6, 1.8 and 2.0 M$_{\odot}$ NS, respectively. While with $\cQMsq/c^2$=1 the $f^{\rm mass}_{\rm QM}$ almost remains the same as in the default calculation where the PDF of $\cQMsq/c^2$ peaks at $\cQMsq/c^2=0.95^{+0.05}_{-0.35}$ at 68\% confidence level as shown in Fig. \ref{QM} (f).

\begin{figure}[htb]
\begin{center}
\hspace{-1.5cm}
    \resizebox{0.55\textwidth}{!}{
 \includegraphics[width=\linewidth]{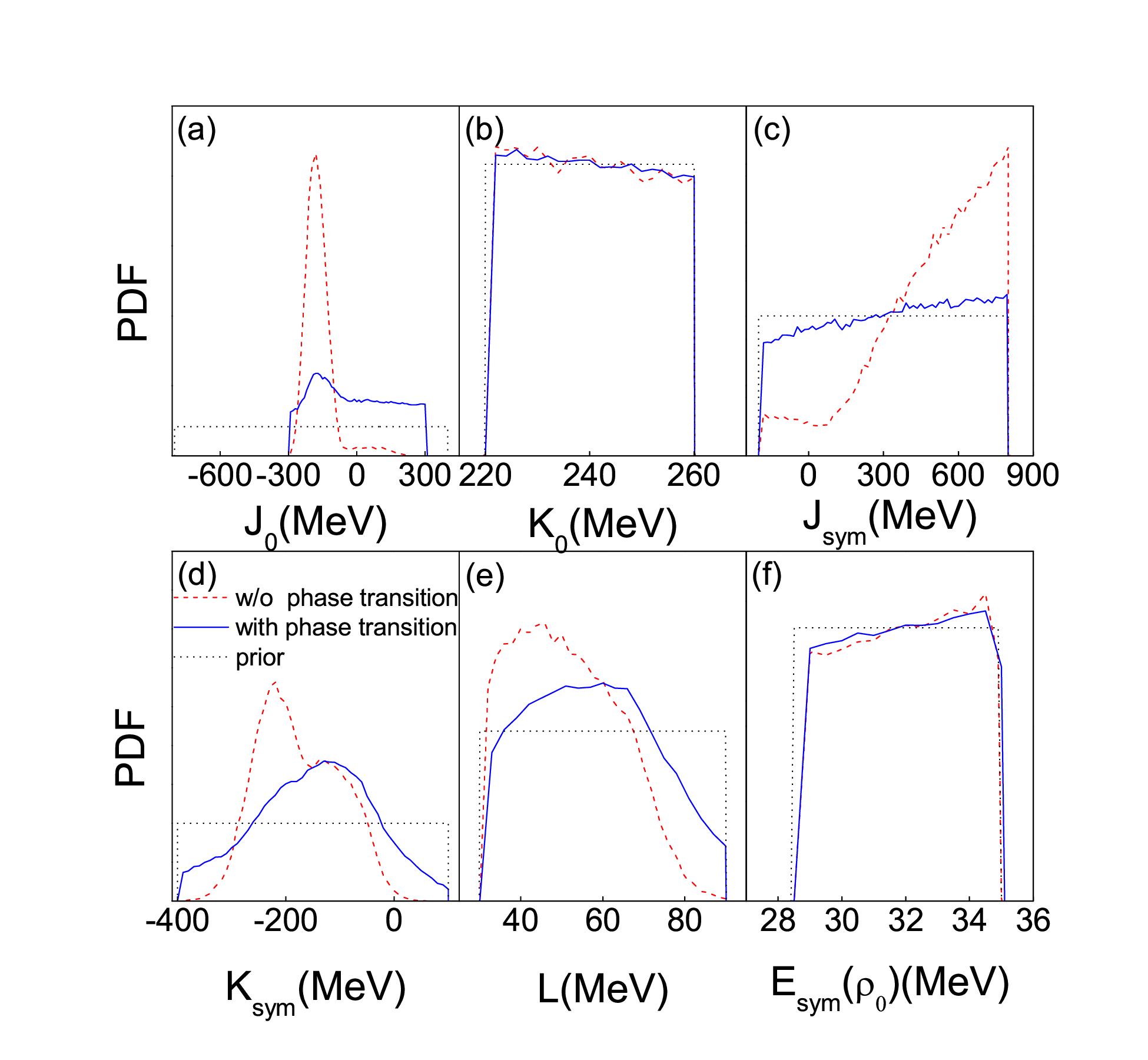}
 }
  \caption{(Color online) The posterior probability distribution functions (in arbitrary units) of nuclear matter EOS parameters inferred from Bayesian analyses with (thick blue curves) and without (thin red curves) considering the hadron-quark phase transition in neutron stars in comparison with their uniform priors (dashed curves). }\label{NMEOS}
\end{center}
\end{figure}
\begin{figure*}[htb]
\begin{center}
 \resizebox{0.9\textwidth}{!}{
  \includegraphics[scale=1]{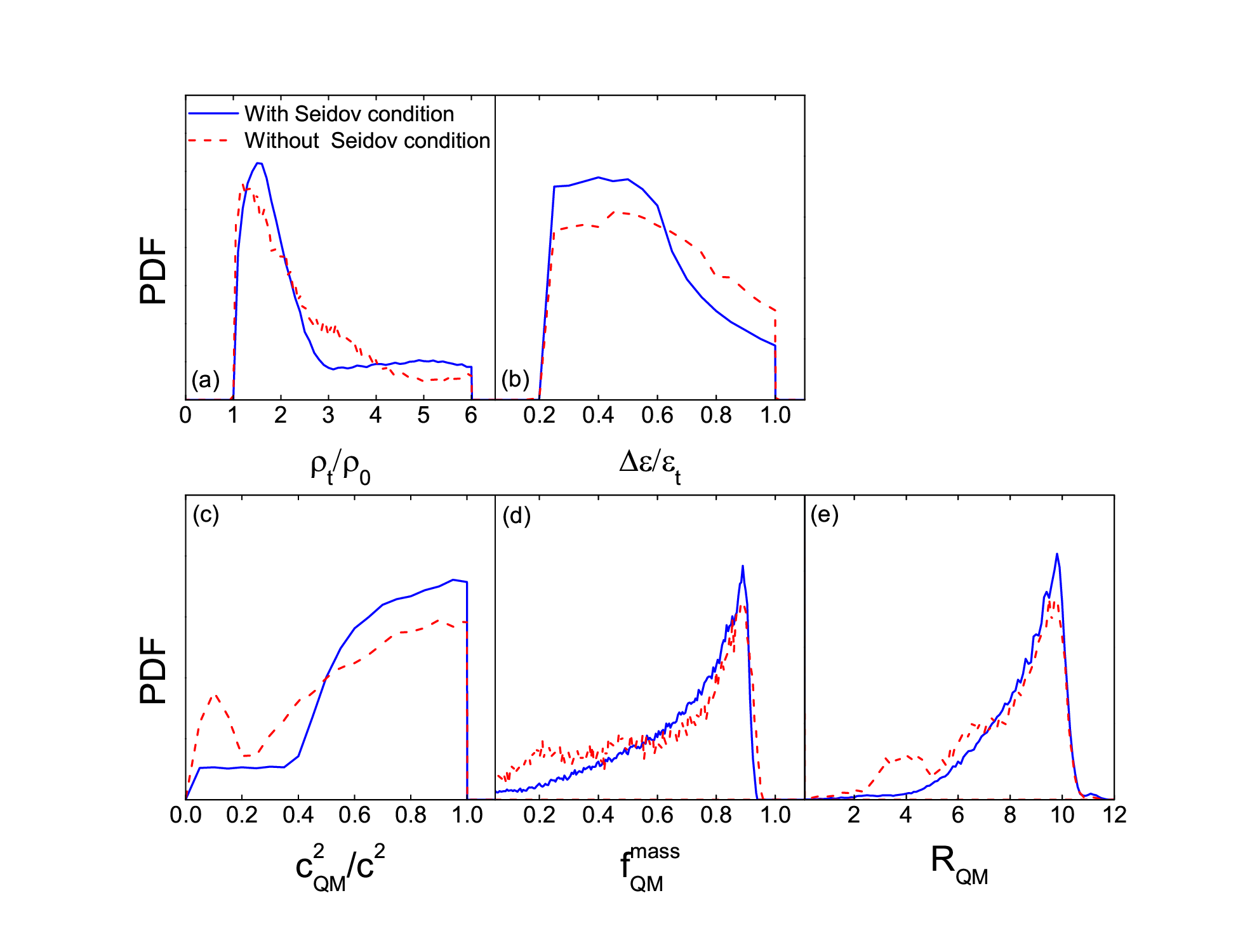}}
\vspace{-0.5cm}
  \caption{(Color online) The posterior probability distribution functions (in arbitrary units) of the three quark matter EOS parameters as well as the fraction and radius of quark matter in hybrid neutron stars of mass 1.4 M$_{\odot}$ with (solid curves) and without (dashed curves) considering the Seidov condition of Eq. (\ref{eqn:stability}).}\label{Seidov}
\end{center}
\end{figure*}

\subsection{Posterior probability distribution functions of nuclear matter EOS parameters extracted with and without considering the hadron-quark phase transition in neutron stars}
Properties of NSs have been studied extensively using various models with or without considering the hadron-quark phase transition in the literature for many years. Within the framework of the present work, it is thus interesting to study effects of considering the hadron-quark phase transitions in NSs on extracting nuclear matter EOSs using astrophysical observables. Shown in Fig. \ref{NMEOS} are our results. Some interesting observations can be made:
\begin{itemize}
\item The incompressibility $K_0$ of symmetric nuclear matter and the symmetry energy $E_{\rm{sym}}(\rho_0)$ at saturation density $\rho_0$ are not affected at all. In fact, their posterior PDFs are not much different from their uniform prior PDFs. These are not surprising and consistent with earlier findings.  While the most probable value of $J_0$ characterizing the stiffness of symmetric nuclear matter at supra-saturation densities does not change, the hadron-quark phase transition requires more contributions from larger $J_0$ values as it generally softens the EOS unless the $\cQMsq/c^2$ is close to 1.

\item The L and $K_{\rm sym}$ parameters together characterize the density dependence of nuclear symmetry energy around $(1-2)\rho_0$. They are known to have significant effects on the radii of canonical NSs in both forward-modelings and Bayesian inferences, see, e.g., Ref \cite{BALI19}, for a recent review. It is seen that their posterior PDFs shift significantly to higher values especially for L when the hadron-quark phase is considered. This can be well understood as the hadron-quark phase transition reduces significantly the pressure above $\rho_t$ compared to the extension of the hadronic pressure into higher density regions. To reproduce the same radius data, the contribution to pressure from the symmetry energy in the $(1-2)\rho_0$ density region has to increase. Thus, the L and $K_{\rm sym}$ parameters have to be higher. Since the $J_{\rm sym}$ characterizes the symmetry energy at densities above about $(2-3)\rho_0$ \cite{Zhang19EPJA}, with the PDF of $\rho_t/\rho_0$ peaks at $1.6^{+1.2}_{-0.4}$ and all the QM EOS parameters are isospin-independent, the analysis considering the hadron-quark phase transition is thus not sensitive to what one uses for the $J_{\rm sym}$. Consequently, the posterior PDF of $J_{\rm sym}$ is almost identical to its prior PDF.  Therefore, the high-density behavior of nuclear symmetry energy extracted from NS properties does depend on whether one considers the hadron-quark phase transition or not. Moreover, the nuclear symmetry energy loses its physical meaning above the hadron-quark transition density.

\item  While the most probable values of L and $K_{\rm sym}$ extracted from the astrophysical data with and without considering the hadron-quark phase transition are significantly different,  they are unfortunately all consistent with
currently known theoretical predictions and findings from terrestrial nuclear experiments \cite{BALI19,NPnews}. Moreover, to our best knowledge, there is currently no terrestrial experimental constraint on the $J_{\rm sym}$ at all. Thus, the available constraints on the nuclear EOS from terrestrial nuclear laboratory experiments do not provide any additional preference on whether QM exists or not in NSs.

\end{itemize}
\subsection{Effects of the Seidov condition on the posterior probability distribution functions of quark matter EOS parameters}\label{Se}
As mentioned earlier, the Seidov stability condition of Eq. (\ref{eqn:stability}) may affect topologies of the mass-radius curve. It may thus also affect the PDFs of quark matter EOS parameters we inferred from the neutron star observables. It is therefore interesting to compare the PDFs of quark matter EOS parameters inferred with and without considering the Seidov stability condition. Shown in Fig. \ref{Seidov} is such a comparison for canonical neutron stars of mass 1.4 M$_{\odot}$. As one expects, by removing the Seidov condition of Eq. (\ref{eqn:stability}) the most probable transition density $\rho_t/\rho_0$ shifts slightly lower from 1.6 to 1.3 while the PDF of the energy density jump has higher weights towards larger $\De\ep/\varepsilon_{t}$ values compared to the default calculation. Interestingly, there are additional bumps around $f^{\rm mass}_{\rm QM}\approx 0.2$ and $R_{\rm QM}\approx 4$ km in the calculation without the Seidov condition besides the major peaks at the same locations as in the default calculation. The minor peaks in the PDFs of  $f^{\rm mass}_{\rm QM}$ and $R_{\rm QM}$ indicate an enhanced formation of a small quark core with a high energy density, probably due to the formation of an additional disconnected stable hybrid branch with smaller radii, compared to the calculation with the Seidov condition.

Since we fixed the neutron star mass at 1.4 M$_{\odot}$, for the nuclear pressure to remain the same but at higher energy densities to reproduce the same set of observables under the same conditions, the speed of sound of quark matter has to become smaller compared to the default calculation. Consequently, the PDF of the speed of sound in quark matter shifts towards lower $\cQMsq/c^2$ values. Interestingly, it has a minor peak around $\cQMsq/c^2=0.1$ that is even less than the 1/3 predicted by the perturbative QCD. It would also be interesting to sort out the posterior events (accepted EOS parameter sets) to study the fractions and topologies of the different branches in the mass-radius plot with and without using the Seidov condition. Such a study within the same Bayesian framework using expected/imagined future radius measurements of more massive neutron stars is in progress and will be reported elsewhere.

\section{Summary and conclusions}
In summary, within the Bayesian statistical framework using generic EOS parameterizations for both the hadronic and quark matter connected through the Maxwell construction we inferred the PDFs of EOS parameters as well as the QM fraction and its size from NS radius data from several recent observations with and without using the Seidov stability condition. We found that the available astrophysical data and all known EOS constraints prefer the formation of a large volume of QM even in canonical NSs regardless whether the Seidov condition is used or not. Future Bayesian inferences using unified EOS models describing both NSs and heavy-ion reactions with possible phase transitions from combined multimessenger data from both fields will significantly improve our knowledge about the EOS of super-dense neutron-rich matter.

\section*{Acknowledgments}
This work was supported in part by the Yuncheng University Research Project under Grant No. YQ-2017005, the U.S. Department of Energy, Office of Science, under Award Number DE-SC0013702, the CUSTIPEN (China-U.S. Theory Institute for Physics with Exotic Nuclei) under the US Department of Energy Grant No. DE-SC0009971, and the National Natural Science Foundation of  China under Grant No. 11505150.


\end{document}